# DO WE KNOW WHAT THE TEMPERATURE IS?

*Jiří J. Mareš*[1]

Institute of Physics of the ASCR, v. v. i., Cukrovarnická 10, 162 00, Prague 6, Czech Republic

**Abstract:** Temperature, the central concept of thermal physics, is one of the most frequently employed physical quantities in common practice. Even though the operative methods of the temperature measurement are described in detail in various practical instructions and textbooks, the rigorous treatment of this concept is almost lacking in the current literature. As a result, the answer to a simple question of "what the temperature is" is by no means trivial and unambiguous. There is especially an appreciable gap between the temperature as introduced in the frame of statistical theory and the only experimentally observable quantity related to this concept, phenomenological temperature. Just the logical and epistemological analysis of the present concept of phenomenological temperature is the kernel of the contribution.

**Keywords:** thermodynamics, temperature scales, hotness, Mach's postulates, Carnot's principle, Kelvin's proposition

**Introduction**

What does the temperature mean? It is a classically simple question astonishingly lacking an appropriate answer. The answers, namely, which can be found in the textbooks on thermodynamics, are often hardly acceptable without serious objections. For illustration, let us give a few typical examples here, the reader can easily find others in the current literature by himself. To the most hand-waving belong the statements such as "the temperature is known from the basic courses of physics" or even "temperature is known intuitively". More frankly sounds the widely used operative definition "the temperature is reading on the scale of thermometer". In contrast to it, rather philosophical is the sentence "temperature is a physical property of a system that underlies the common notions of hot and cold." which, nevertheless, rather specifies what the temperature should be, giving no idea of what it actually is. To high scientific standard pretends the definition "on the macroscopic scale the temperature is the unique physical property that determines the direction of heat flow between two objects placed in thermal contact", being in fact an explanation "*obscurum per obscurius*" which transforms the temperature problem to the problem of flow of something more uncertain. Real nightmare is then for students statement "the absolute temperature is integrating divisor converting heat (imperfect differential) into an exact differential".

Of course, there is no doubt that the temperature is a central concept of thermal physics and that is why a lot of researchers were trying, in different ways, to bring the temperature concept on the safer grounds. For example, the authors preferring the axiomatic approach are, as a rule, inclined to assume that the temperature is a primitive concept which need not be, in principle, derived from other presumable more primitive ideas. Unfortunately, the experimental determination of temperature in any particular case requires performing of a lot of non-trivial operations which should be substantiated by its definition. Thus such a shift from an operative physical definition to a metaphysical one, very satisfactory for theoreticians, makes any actual temperature measurement performed by experimentalists an inexplicable obscure ritual. For researchers who consider the thermal physics being nothing but an outgrowth of statistical mechanics, the statistical temperature *T* of a system can be defined by a formula [1,2]

$$1/T = k \, (d \ln \Omega / dE), \qquad (1)$$

where $\Omega$ is the number of equally likely configurations (microstates) of the system, *E* its energy and $k = 1.38 \times 10^{-23}$ JK$^{-1}$ the Boltzmann constant. Since, however, practical measurements of temperature are made by macroscopic thermometers and not by means of evaluation of statistical data related to aggregates of particles and excitations existing in a given body, it is apparent that statistical definition of temperature cannot serve as a full substitute for the phenomenological one. Moreover, the experimental evidence for identification of both temperatures, phenomenological and statistical, is till now incomplete and even controversial.

---

[1] maresjj@fzu.cz

In this contribution we critically analyse the present concept of phenomenological temperature putting emphasis on the relevant experimental aspects and anthropic elements involved, partially following the historical development of this fundamental physical entity.

**Thermoscope**

The first step toward the establishment of temperature concept was very likely the rediscovery of correlation between human sensations, represented by a series cold, cool, tepid, warm, hot, and observable physical state of sometimes very curious devices described in the first Latin translation of Hero's "Pneumatica"(1575) [3]. Substitution of some device, "*thermoscope*", for human body then started *objectification process* resulting eventually in a definition of a new physical quantity, temperature. It was clear very early that the key for the quantification of thermal phenomena was just the understanding to these devices. Skipping now the otherwise interesting history [4] of experimentation with various thermoscopes and thermometers, we turn our attention rather to their modern description.

Adequate for a systematic approach to the thermoscopic devices is a theory of homogeneous two-parameter systems. Generally, the state of any homogeneous body is described by a system of empirical constitutive relations written in terms of suitably chosen parameters. The number of parameters and constitutive relations can be, however, according to so called Fürth's conjecture [5], step-wise reduced by keeping arbitrarily chosen parameters constant. As a result, the body is described completely by only two parameters coupled by a single constitutive relation, provided that the external conditions remain constant. As can be shown, it is, moreover, possible to choose the couple of two remnant parameters $X$, $Y$, in such a way that it satisfies the following dimensional relation (square brackets mean here the physical dimension of the quantity enclosed)

$$[\text{Energy}] = [X] \times [Y] \qquad (2)$$

in which one of the quantities, say $X$, is intensive while the other one, $Y$, is extensive [6]. Such a couple of quantities obeying relation (2) is then called a couple of *conjugate variables*. For example, in mechanics the role of conjugate variables play a generalized coordinate and a generalized force, the product of which constitutes the term entering the energy balance equation. The existence of the intensive and extensive "aspects" of heat which was already recognized by J. Black [7] is thus in this context the discovery of primary importance for the formalization of theory of heat and its compatibility with other branches of physics. His "intensity of heat" and "matter of heat" can be, namely, quite naturally assigned to a certain couple of conjugate variables, tentatively called "temperature" and "heat".

The treatment of real systems in terms of conjugate variables enables one to introduce fundamental concepts of thermal physics without *a-priori* reference to thermal phenomena *per se*. A primary role plays here the procedure used in practical thermometry for proving "thermal contact", so called *correlation test* which provides simultaneously the basis for an operational definition of diathermic and adiabatic partitions [8], viz: Let us have two systems characterized by two couples of conjugate variables $(X,Y)$ and $(X',Y')$, respectively, which are separated by a firm partition. The partition is then called *diathermic* if the changes of $(X,Y)$ induce changes of $(X',Y')$ and vice versa (positive result of correlation test). In the case of a negative result of the correlation test the partition is called *adiabatic*. These definitions enable one also to extend the concept of equilibrium to the region of thermal phenomena. Indeed, taking into account the standard formulation for the equilibrium state of a two-parameter system: "Any state of a body in which conjugate variables remain constant so long as the external conditions remain unchanged is called equilibrium state" and adding then the concept of diathermic partition, we obtain a definition

(i) *If two bodies being separated by a diathermic partition are both in equilibrium state, they are in thermal equilibrium.*

In the frame of the theory of two-parameter systems a thermoscope can be then formally introduced by the following definition (cf. [9]):

(ii) *Thermoscope is any two-parameter system in which one of the conjugate variables, say Y, is fixed ($Y=Y_0$). It is further assumed that the thermoscope can be brought into diathermic contact with other bodies and that it is sufficiently small in comparison with these bodies, in order not to disturb their thermal equilibrium. The second conjugate parameter is then called a thermoscopic variable.*

**Thermoscopic states**

Notice that in definition (ii) we have in fact applied Fürth's parameter reduction to the two-parameter system. For resulting single-parameter system at constant external conditions, however, *X* should be constant. Any change of thermoscopic variable *X* thus reveals a change of external conditions vicinal to thermoscope. Since such a change is not due to the change of parameters which are under our control, it is assumed that it is a result of till now unspecified thermal effects which can be observed just in this way. These important facts are sometimes referred to as the *residual nature of thermal effects* [10].

Thermoscopic variables are generally of quite a diverse physical nature. Length, volume, resistance, voltage, frequency and many others may be chosen for variable *X*. (For the sake of brevity we are not distinguishing in this paper symbols for physical quantity and its numerical value.) To distinguish among various thermoscopic variables, different thermoscopes and physical conditions under which they operate small Latin indexes are used. According to this convention, reading $X_k(P)$ of *k*-th thermoscope which is in diathermic contact with a body under investigation corresponds to the *thermoscopic state P* of the body.

The whole set of the thermoscopic states which can be observed in this way is then marked as $\mathbf{H}_k$. Notice that the readings $X_k$ are related to the thermoscope, while the indicated thermoscopic state such as $P \in \mathbf{H}_k$ already relates to the body. Obviously, the thermoscopic state $P$ can be shared by other bodies in the Universe, e.g. by those for which the reading on the $k$-th thermoscope, being with them in diathermic contact, is just $X_k(P)$.

Taking now into account some requirements on the mathematical structure of thermoscopic variables $X_k$, important properties of sets $\mathbf{H}_k$ may be found. For example, it can be assumed that thermoscopic variables are commensurable quantities (Cf. Mareš JJ. invited lecture: Mathematical Structure of Physical Quantities. 4th IC- FQMT Prague 2013), in contrast to our previous work [11], where thermoscopic variables were considered to be real, i.e. that $X_k \in \mathbf{I}_k$, for every index $k$, where $\mathbf{I}_k$ means a certain segment of rational numbers ($\mathbf{I}_k \subset \mathbf{Q}$, $X_k \in \mathbf{Q}$). If then for every couple of thermoscopic states $P, Q \in \mathbf{H}_k$

$$P \neq Q, \quad \Leftrightarrow \quad X_k(P) \neq X_k(Q), \tag{3}$$

the set $\mathbf{H}_k$ can be ordered ($\prec, \succ$) in accordance with an intrinsic order already existing in the rational segment $\mathbf{I}_k$, ($<, >$) respecting the following equivalences:

$$\begin{aligned} P \prec Q &\Leftrightarrow X_k(P) < X_k(Q) \\ P \succ Q &\Leftrightarrow X_k(P) > X_k(Q) \\ P = Q &\Leftrightarrow X_k(P) = X_k(Q). \end{aligned} \tag{4}$$

Of course, we have here a liberty to choose the "arrow" of ordering, substituting the symbols $<, >$ for $>, <$. Relation (4) ensures simultaneously that the topological features of rational segment $\mathbf{I}_k$ are preserved also in the set $\mathbf{H}_k$, namely, that $\mathbf{H}_k$ must be countable and dense in itself [12].

There is another, quite a natural physical requirement which guarantees the consistency of the concept of thermoscopic states known as a *principle of indifference* [10]:

(iii) *Different thermoscopes operating in the common range of thermoscopic states should distinguish any two different states $P \neq Q$, $P, Q \in (\mathbf{H}_k \cap \mathbf{H}_j)$, regardless of their constructions, thermometric substances and thermoscopic variables used.*

The validity of this principle can be in any particular case tested experimentally constructing the so called Dulong-Petit plot for two thermoscopes $k$ and $j$ [13]. It is a locus of readings $X_k$ versus $X_j$ if both thermoscopes are kept in diathermic contact with the same thermal bath. Obviously, if such a plot is in a certain range monotonic, the thermoscopes satisfy there the principle of indifference. A thermoscope using as thermometric fluid pure water may serve as a good example of application of the principle of indifference. The Dulong-Petit plot with respect to practically all other thermometers reveals there, due to the well-known water anomaly [14], in the neighborhood of ~ 4°C non-monotonic behavior. However, in the ranges between say 1°C - 4°C and between 4°C - 100°C the principle of indifference is valid, so that the water dilatometer can be in these ranges used as a regular thermoscopic device.

**Fixed points, Mach's postulates**

A serious obstacle for the development of thermometry was an appreciable irreproducibility of early thermoscopes. The attempts to solve this problem by making exact copies of a standard instrument were only partially successful [15]. An important qualitative step toward the scientific thermometry was therefore the discovery and general use of *fixed thermometric points* (shortly fixed points) serving as a fiducial points for thermoscopes of any kind [16]. (As examples of fixed points may serve melting or boiling points of water, boiling point of helium, melting point of platinum, etc., all at normal atmospheric pressure). Indeed, the use of various fixed points, the common properties of which are specified bellow, increased the reproducibility of readings of thermometric devices appreciably.

(iv) *There exist in definite way prepared bodies called fixed (thermometric) points, which unambiguously define certain thermoscopic states, i.e. enable their identification or reestablishment. The fixed points constitute a set of fixed points $\in \mathbf{F}$.*

As was recognized just before the end of nineteenth century the significance of fixed points is not confined only to a calibration of thermoscopes but that it has a primary importance also for the theory of thermometry. We have to mention here especially two facts neglected for a long time, namely, that it is always possible to find in an operation range of any thermoscope a sufficient number of fixed points enabling its calibration and that it is always possible to find out fixed points outside of any interval of thermoscopic states.

Apparently, there are no principal but only technical limits on preparation of new fixed points. Generalization of these empirical observations by means of incomplete induction is due to E. Mach who formulated the following *Mach's postulates* [13]:

(v)  *1) Fixed points can be ordered ($\succ,\prec$)*

  *2) To every fixed point can be ever found a fixed point which is lower ($\prec$) or higher ($\succ$)*

  *3) An inter-lying fixed point can be ever constructed*

In terms of mathematical set theory [12] these three Mach's postulates can be put into a more condensed form:

(vi) *The set of fixed points **F** is an infinite countable ordered dense set having no first and no last point.*

It should be stressed here that veracity of Mach's postulates was never disproved by experiment. Speaking for a while in terms of Kelvin's temperature, the temperatures observed range from ~$10^{-10}$ K (Low Temperature Lab, Helsinki University of Technology) up to ~$10^9$ K (supernova explosion) without any traces that the ultimate limits were actually reached. Speculative upper limit provides only the so called Planck temperature $T_P = \sqrt{(\hbar c/G)} \times (c^2/k) \approx 1.417 \times 10^{32}$ K, hypothetically corresponding to the first instant of the Big Bang and depending on the assumption that the constants c, G and k involved (speed of light, gravitational constant and Boltzmann's constant) are really universal. Therefore the conjectures involved in Mach's postulates, i.e. that the set **F** and consequently hotness series **H** (see below) has no highest and no lowest point, is obviously operating at least for all thermal phenomena already known.

**Hotness series, empirical temperature scale**

Since, as we have seen above, for any fixed point $\Pi \in \mathbf{F}$ there exists a point $P \in \mathbf{H}_k$, it means that the set F can be ordered by means of essentially the same relation ($\succ,\prec$) as $\mathbf{H}_k$, (first Mach's postulate). Giving to these statements a physical meaning, we can say that the calibration of thermoscopes using fixed points can be interpreted as an ordering of set **F**. On the other hand, fixed points are very useful for sewing-up together overlapping sets of thermoscopic states, $\mathbf{H}_k$. In order to cover much larger range of thermoscopic states it is, namely, necessary to combine the thermoscopes of different construction, very often working with different thermoscopic variables. Operational method for sewing-up together overlapping sets of thermoscopic states can be described as follows (cf. [17]).

Let us assume that two sets of thermoscopic states overlap i.e. that $\mathbf{H}_k \cap \mathbf{H}_{k+1} \neq \varnothing$. In order to realize this fact in experiment one has to find a fixed point $\Pi \in \mathbf{F}$ for which the corresponding thermoscopic state P belongs to both $\mathbf{H}_k$ and $\mathbf{H}_{k+1}$, i.e. $P \in \mathbf{H}_k$, $\mathbf{H}_{k+1}$. Without loss of generality we can further construct the subsets $\mathbf{H'}_k \subset \mathbf{H}_k$ and $\mathbf{H'}_{k+1} \subset \mathbf{H}_{k+1}$ in such a way that $Q \prec P$ for every $Q \in \mathbf{H'}_k$ and $R \succ P$ for every $R \in \mathbf{H'}_{k+1}$. Obviously, ordering of thermoscopic states in sets $\mathbf{H'}_k \cup \mathbf{H'}_{k+1} = \mathbf{H}_k \cup \mathbf{H}_{k+1}$ will correspond below $P$ to the ordering in $\mathbf{H}_k$ and above $P$ to that in $\mathbf{H}_{k+1}$. Applying repeatedly the procedure just described and simultaneously looking for new fixed points and for new physical effects enabling the construction of new kinds of thermoscopic devices, we can built a chain of $\mathbf{H}_k$'s more and more extending the region of accessible thermoscopic range. Sharing then the belief of professor Mach that such a procedure is limited only by our skills, we can assume that it is possible to construct the set involving all thinkable thermoscopic states,

$$\mathbf{H} = \cup_k \mathbf{H}_k, \qquad (5)$$

which is called *hotness series* and the elements of which are called *hotness levels* (renamed thermoscopic states). As for topological properties, the hotness series, being a countable union of countable sets $\mathbf{H}_k$ (notice, the number of thermoscopes is countable), is also countable. Moreover, because the cardinality of both sets $\mathbf{H}$ and $\mathbf{F}$ is the same, these sets are equivalent ($\mathbf{F} \Leftrightarrow \mathbf{H}$) and due to the fact that the order in $\mathbf{F}$ is induced by the order in $\mathbf{H}$ ($\succ, \prec$), one-to-one order–preserving mapping can be established between $\mathbf{F}$ and $\mathbf{H}$.

Putting now all these facts together, we are in position to define e*mpirical temperature scale,* which is a projection of an abstract set, hotness series $\mathbf{H}$, representing in a philosophical sense something as a "platonic idea of temperature", on the set of numbers.

(vii) *Empirical temperature scale $\theta$ is any order preserving one-to-one mapping of hotness series* $\mathbf{H}$ *on a segment of rational numbers* ($\theta \in \mathbf{Q}$).

An empirical temperature scale $\theta$ just defined enables one to reestablished hotness levels $\in \mathbf{H}$ and even perform some scientific measurements; however, it does not represent actual physical quantity which is necessary for theoretical treatment of effects observed.

Moreover, empirical temperature scales being only order-invariant, provide an enormous room for anthropic constructions. To convert such arbitrary empirical scales into the scales defining a regular physical quantity, temperature, it is necessary to take into account some reasonably chosen auxiliary criteria. There are essentially four types of scales assigning the numbers to entities enabling their quantification in this way. These scales are classified according to their invariance with respect to various mathematical operations [18]. One can thus distinguish nominal, order, interval and ratio scales, which are invariant to permutation, order-preserving, linear and similarity group of transformations, respectively. In physics, in contrast to e.g. psychology, we are as a rule satisfied only with the most perfect type of the scale, *ratio scale*, which is the only one generating *physical quantity*. The auxiliary condition we are looking for thus should select from the whole class of empirical scales its subclass, which would be invariant with respect to similarity transformations, i.e.

$$\theta = \alpha\, \theta'. \tag{6}$$

In other words, all the alternative temperature scales defining the temperature being a physical quantity can therefore differ only by a constant similarity factor $\alpha$. Inherent property of this group satisfying condition (6) is the existence of common limit for $\theta \to 0$. This common limit performs not a common value but it is a common greatest lower bound for values of all $\theta$'s. Indeed, according to second Mach's postulate, **F** and consequently **H** have no lowest points, i.e., there cannot be such a hotness level $P \in \mathbf{H}$ for which $\theta(P) = 0$. Accepting thus Mach's postulates and the requirement that the temperature is a regular physical quantity the law of *unattainability of absolute zero temperature* [19] must be satisfied automatically, without problematic "proofs" which can be currently found in the literature [20].

**Carnot's principle and Kelvin's proposition**

Among various auxiliary criteria applicable to the conversion of empirical scales into ratio scales broke through two idealized anthropic models of reality, namely, *ideal heat engine* and *ideal (perfect) gas*. The first idealization took its origin in early experiments on the development of mechanical work by means of heat engines. In spite of the fact that these experiments were rather primitive and of doubtful accuracy, their analysis enabled S. Carnot to introduce some important theoretical concepts, such as the *cyclic process* for periodically working engine and *reversible process* for the case where the heat engine works without wastes and generation of heat. For heat engines utilizing then the cyclic reversible process (so called *ideal heat engines*) Carnot was able to formulate the following theorem ("principle") which in its original version reads [21]

(viii) *The motive power of heat is independent of the agents set at work to realize it; its quantity is fixed solely by the temperatures of the bodies between which, in final result, the transfer of the caloric occurs.*

Regardless of its archaic form and of the use of different terms for the apparently same thing (heat, caloric) the principle can be mathematically formulated as

$$L = \varsigma\ F(\theta_1, \theta_2), \quad (7)$$

where $L$ is the motive power or work done, $\varsigma$ is the non-specified extensive quantity representing reversibly transferred heat and $\theta_1$, $\theta_2$ are the empirical temperatures of heater and cooler, respectively. Formula (7) may be rewritten also in the form emphasizing the role of difference of empirical temperatures

$$L = \varsigma\ F'(\theta_1)\ (\theta_1 - \theta_2), \quad (8)$$

where the correction function $F'(\theta)$ is called Carnot's function [22]. Using for the cooler of heat engine e.g. a bath with melting ice or with boiling liquid helium, this bath can serve simultaneously as a fixed point keeping the temperature $\theta_2$ and as a mean for measuring the transferred heat $\varsigma$. (Heat is then measured by an amount of melted ice or of evaporated helium.) Changing heater temperature, Carnot's function can be determined experimentally from relation (8) as a function of $\theta_1$. At this stage a revolutionary step toward the definition of temperature scale independent of particular type of thermometer and thermometric substance was made by Lord Kelvin [23]. He proposed to treat Carnot's theorem not as a heuristic statement deduced from experiments of rather a limited accuracy but as a fundamental principle of absolute validity (*Kelvin's proposition*). Since accordingly, Carnot's function must be the same for all substances; it can depend only on the empirical temperature scale $\theta$ used. *Mutatis mutandis*, prescribing then a convenient analytical form to Carnot's function instead of determining it experimentally, an "absolute" (substance and device independent) temperature scale will be unambiguously defined. Giving to Carnot's function the simplest permissible analytical form

$$F'(\theta_1) = 1, \tag{9}$$

we obtain from (8) a relation defining Kelvin's temperature scale ($\theta_1$ and $\theta_2$ are renamed here to $T_1$ and $T_2$ respectively)

$$L/\varsigma = \beta (T_1 - T_2), \tag{10}$$

where the numerical factor β depends only on the system of units chosen for the measurement of work and heat. Remarkably, the definition proposed is based on the caloric theory of heat and on the acceptance of prescription (9) called sometimes "caloric gauge" [24]. According to this theory the temperature (intensive aspect of heat) plays the role of potential [25] of substance-like entity called caloric (extensive aspect of heat) and the corresponding quantities $T$ and $\varsigma$ make up a couple of conjugate variables obeying dimensional equation quite analogous to (2) i.e.

$$[Energy] = [T] \times [\varsigma]. \tag{11}$$

Similar approach to the definition of temperature in the frame of dynamical theory of heat, thermodynamics, is rather difficult. According to Joule's "principle of equivalence of energy and heat" [26,27], namely,

$$L/\varsigma = J, \tag{12}$$

where the dimensionless constant J is known as a mechanical equivalent of heat (J = 4.18 J cal$^{-1}$). Identification of heat with a special form of energy, makes the term $F'(\theta_1)(\theta_1 - \theta_2)$ in (8) constant, what effectively separates temperature from heat. Moreover, temperature and heat are no more conjugate variables (Joule ≠ Kelvin × Joule) and the "thermal term" in the energy balance equation must be written using a somewhat artificial quantity, entropy [28], of not very clear phenomenological meaning. As a result, in thermodynamics, we have for extensive aspect of heat instead of one, two extensive quantities, heat (= form of energy) and entropy [29]. Joule's arbitrary postulate, "the principle of equivalence", sometimes proclaimed to be one of the "greatest achievements of experimental science" [30] which it is not, thus enormously complicates not only the introduction of temperature but also the conceptual basis of thermal physics as a whole [31]. The author is convinced that just this fact is responsible for rather a poor understanding of the concept of temperature as has been mentioned in the introduction.

**Ideal gas scale**

An alternative approach to the construction of temperature scale ($T°$) is based on the hypothetical substance known as *ideal (perfect) gas*. This idealization generalizes the most salient common features of real rarified gases where the long-range interaction between gas molecules is reduced. The constitutive relation controlling the behavior of the ideal gas thus reads

$$T° = pV / nR, \qquad (13)$$

where $p$ and $V$ are respectively the pressure and the volume of n moles of ideal gas closed in the thermometer bulb and R is the gas constant (R = 8.3145 JK$^{-1}$mol$^{-1}$). Formula (13) reveals remarkable symmetry with respect to quantities $p$ and $V$, so that we can exploit anyone of these two quantities as a thermoscopic variable keeping the other constant. Comparing these two cases, inevitably:

$$T°_p = T°_V = T°, \qquad (14)$$

where $T°_p$ and $T°_V$ are temperatures determined by means of constant pressure and constant volume method, respectively. The exact realization of condition (14) in experiments with real gases and with prescribed high accuracy (typically of order 0.1%) is a very difficult task. However, Berthelot [32] devised a simple graphical method which enabled one to extrapolate experimental data obtained in real gases at finite pressures to the case corresponding to the ideal gas and finally determined also the value of $T°$ satisfying conditions (14).

Let us now compare the properties of empirical scale (13) with Kelvin's scale as defined by relation (10). For this purpose we will make a thought experiment with gas thermometer filled with an ideal gas and treated as an ideal heat engine. Let us realize a reversible cycle C consisting of two isothermic processes at temperatures $T°_1$ and $T°_2$ completed by two isochoric processes at volumes $V_1$ and $V_2$, respectively. The work done will be according to equation (13)

$$L = \oint_C p\,dV = (T°_1 - T°_2)\, nR\, \ln(V_2/V_1). \tag{15}$$

Identifying the heat $\varsigma$ transfered during the cycle via the ideal gas from heater to cooler with the second term on the right side of equation (15), i.e.

$$\varsigma = nR\, \ln(V_2/V_1), \tag{16}$$

we can rewrite equation (15) in the form

$$L/\varsigma = (T°_1 - T°_2). \tag{17}$$

Evidently, this formula is fully congruent with equation (10) defining Kelvin's temperature scale. Moreover, by a proper choice of constant factors involved, it is possible to transform linear relations (10) and (17) into the same ratio scale, $T° = T$. Such a reduction of number of arbitrary parameters may be effectively achieved by *calibration*. For properly calibrated scales $T°$ and $T$, the following theorem can be then formulated:

(ix) M*easurement of temperatures by means of ideal gas thermometer is equivalent to the measurement of temperatures by means of ideal heat engine.*

This theorem is of primary importance in thermal physics, both experimental and theoretical. While Kelvin's scale based on reversible cycle of ideal heat engine enables one to perform thought experiments in energy representation, the ideal gas scale imitating in fact the behavior of real gases is closer to the experiment and to the molecular model of matter. Theorem thus serves as a bridge interconnecting epistemologically different concepts of thermal physics.

**Calibration of Kelvin's scale**

In order to finish establishment of temperature ratio scale defining the physical quantity, temperature, it is thus necessary to perform a calibration procedure.

For calibration of thermometers and definition of empirical temperature scales two alternative methods were traditionally used. The first one, independently proposed by Boyle, Hooke and Huygens in second half of the 17th century [15,33], each mark on the scale of thermoscope corresponded to an expansion or contraction about 1/1000 of the volume of the thermometric fluid at an ice point (fixed point, where the water just begins to freeze at normal pressure). According to the second method, widely used by e.g. Roemer, Fahrenheit, Réaumur and Celsius [34], the scale of thermometer is calibrated by dividing the interval between marks corresponding to two fixed points into equal parts. The main disadvantage of both these methods, i.e. the dependence on the properties of thermometric substance and device used, was, as we saw above, successfully eliminated by Kelvin's construction. For establishment of Kelvin's ratio scale represented by one of the straight lines of group (6) it is thus sufficient to choose only one fixed point (the second one is the common "unattainable" zero), while no reference to thermometric substance is required. At present it is instead of ice point as a fiducial fixed point used the triple point of water [35,36] which was recognized to be the most accurate temperature standard available, independent of external conditions [37,38]. The triple point of water is realized by a quartz cell containing pure air-free water with well defined isotopic composition (corresponding to the Viena Standard Mean Ocean Water) enabling coexistence of all three phases of water, i.e. of ice, liquid water and water vapor. The nominal value of this fixed point being 273.16 K exactly has precision (reproducibility) better then < 0.1 mK and drift < 0.01 mK/year [35].

**Conclusions**

Summarizing, in this paper we have shown that the present central concept of thermal physics is based on two experimentally accessible entities, a set of fixed points **F** and a hotness series **H,** which both are mathematically ordered countable sets dense in themselves. Any empirical temperature scale is then defined by otherwise arbitrary one-to-one order preserving mapping on some rational segment. From the whole system of possible empirical temperature scales a subset of temperature scales for which the empirical temperature has properties of a regular physical quantity is selected by means of two idealized anthropic models, namely, ideal heat engine and ideal gas. Besides the epistemological aspects of our analysis of the concept of temperature, the knowledge of logical structure of the introduction of this important physical quantity, which is only partially discussed in the current literature, may have also a practical impact. It enables one e.g. to extend considerations on essentially non-equilibrium situations [39,40] and to suggest the solutions of some difficult problems in relativistic thermodynamics [41] or in solid-state physics [42].

**References**

[1] Landau L, Lifshitz E. Statisticheskaya fizika. Leningrad: GITTL;1951.

[2] Blundell SJ, Blundell KM. Concepts in thermal physics. New York: Oxford University Press, Inc.; 2008.

[3] Boas M. Hero's Pneumatica: A Study of Its Transmission and Influence. Isis 1949; 40: 38-48.

[4] Mareš JJ. On the Development of the Temperature Concept. J. Therm. Anal. Calor. 2000; 60: 1081-1091.

[5] Fürth R. Algemeine Grundlagen der Physik, Prinzipien der Statistik. In: Handbuch der Physik Bd. IV. Berlin: Springer; 1929.

[ 6] J. Palacios J. Dimensional Analysis. London: Macmillan & Co. Ltd.;1964.

[7] Black J. Lectures on the Elements of Chemistry. Edinburgh: W. Creech; 1803, German translation: Vorlesungen über die Grundlehren der Chemie. Hamburg: Crell; 1804.

[8] Carathéodory C. Untersuchungen über die Grundlagen der Thermodynamik. Math. Ann. 1909; 67: 355-386.

[9] Balamuth L, Wolfe HC, Zemansky MW. The Temperature Concept from the Macroscopic Point of View. Am. J. Phys. 1941; 9: 199-203.

[10] Barnett MK. The Development of Thermometry and the Temperature Concept. Osiris 1956; 12: 269- 341.

[11] Mareš JJ. Hotness Manifold, Phenomenological Temperature. In: Glassy, Amorphous and Nano-Crystalline Materials. Dordrecht: Springer; 2011. pp. 327-346.

[12] Huntington EV. The Continuum and Other Types of Serial Order. New York: Harvard University Press; 1917. Reprint. New York: Dover Phoenix Editions; 2003.

[13] Mach E. Die Principien der Wärmelehre. Leipzig: Verlag von J. A. Barth; 1896.

[14] Epstein PS. Textbook of Thermodynamics. New York: J. Wiley and Sons;1954.

[15] Boyer CB. Early Principles in the Calibration of Thermometers. Am. J. Phys. 1942; 10: 176-180.

[16] Hoppe E. Geschichte der Physik. Braunschweig: Vieweg und Sohn, a. G.; 1926. p. 170.

[17] Serrin J. The Concepts of Thermodynamics. In: Contemporary Developments in Continuum Mechanics. Amsterdam: North-Holland Publ. Co.; 1978. pp. 411-451.

[18] Stevens SS. On the theory of scales of measurement. Science 1946; 103: 677 – 680.

[19] Nernst W. The New Heat Theorem. Reprint: New York: Dover Publications, Inc.; 1969.

[20] Boas ML. A Point of Logic. Am. J. Phys. 1960; 28: 675-675.

[21] Carnot S. Réflexions sur la puissance motrice du feu et sur les machines propres à développer cette puissance. Paris: Bachelier; 1824. German transl.: Ostwald's Klassiker. Nr. **37**. Leipzig: Engelmann; 1909.

[22] Callendar HL. The Caloric Theory of Heat and Carnot's Principle. Proc. Phys. Soc. London 1911; 23: 153-189.

[23] Thomson W (Lord Kelvin of Largs). On the Absolute Thermometric Scale founded on Carnot's Theory of the Motive Power of Heat. Phil. Mag. 1848; 33: 313-317.

[24] Mareš JJ, Hubík P, Šesták J, Špička V, Krištofik J, Stávek J. Phenomenological approach to the caloric theory of heat. Thermochimica Acta 2008; 474: 16-24.

[25] Fuchs HU. The Dynamics of Heat. New York: Springer; 2010.

[26] Joule JP. New Determination of the Mechanical Equivalent of Heat. Phil. Trans.Roy. Soc. London 1878; 169: 365-383.

[27] Bailyn M. A Survey of Thermodynamics. New York: AIP Press; 1990.

[28] Clausius R. Mechanische Warmetheorie. Braunschweig: Viewg Sohn;1876.

[29] Job G. Neudarstellung der Wärmelehre – Die Entropie als Wärme. Frankfurt am Main: Akad. Verlagsges.; 1972.

[30] Shamos MH. Great Experiments in Physics. New York: Dover Publications, Inc.; 1953.

[31] Job G, Lankau T. How Harmful is the Firt Law? Ann. N.Y. Acad. Sci. 2003; 988: 171-181.

[32] Wensel, HT. Temperature and Temperature Scales. J. Appl. Phys. 1940; 11: 373-387.

[33] Burckhardt F. Die Erfindung des Thermometers und seine Gestaltung im XVII. Jahrhundert. Basel; 1867.

[34] von Oettingen AJ. Abhandlungen über Thermometrie von Fahrenheit, Réaumur, Celsius. Ostwald's Klassiker No.57. Leipzig: W. Engelmann; 1894.

[35] ITS-90. Supplementary Information for the International Temperature Scale of 1990. BIPM, ISBN 92-822-2111-3.

[36] Preston-Thomas, H. The International Temperature Scale of 1990 (ITS 90). Metrologia 1990; 27: Z186-Z193.

[37] Moser H. Der Triplepunkt des Wassers als Fixpunkt der Temperaturskala. Ann. d. Phys. 1929; 393: 341-360.

[38] Stimson HF. The Measurement of Some Thermal Properties of Water.

J. of Washington Acad. Sci. 1945; 35: 201-218.

[39] Šesták J. Is the original Kissinger equation obsolete today - not obsolete the entire non-isothermal kinetics? J Therm Anal Calorim. 2014; 117: 3-7.

[40] Šesták J. Kinetic phase diagrams as a consequence of sudden changing temperature or particle size. J Therm Anal Calorim. 2015; doi: 10.1007/s10973-014-4352-8

[41] Mareš JJ, Hubík P, Šesták J, Špička V, Krištofik J, Stávek J. Relativistic transformation of temperature and Mosengeil-Ott's antinomy. Physica E 2010; 42: 484-487.

[42] Mareš JJ, Hubík P, Krištofik J, Nesládek M. Selected topics related to the transport and superconductivity in boron-doped diamond. Sci. Technol. Adv. Mater. 2008; 9: 044101-044107.